\documentclass{article}
\usepackage[dvips]{graphicx}

\title{Measurement of $\beta\beta$ Decay-Simulating Events in Nuclear Emulsion 
with Molybdenum Filling}

\author{V.D.~Ashitkov$^1$, A.S.~Barabash$^1$, V.Ya.~Bradnova$^2$, V.A.~Ditlov$^1$,\\ V.V.~Dubinina$^1$, N.P.~Egorenkova$^1$, S.I.~Konovalov$^1$, E.A.~Pozharova$^1$,\\ N.G.~Polukhina$^3$, V.A.~Smirnitsky$^1$, N.I.~Starkov$^3$, M.M.~Chernyavsky$^3$,\\
T.V.~Shchedrina$^3$ and V.I.~Umatov$^1$\\ [0.4cm]
$^1${\it\small Institute of Theoretical and Experimental Physics,}\\
{\it\small B.~Cheremushkinskaya 25, 117259 Moscow, Russia} \\
$^2${\it\small Joint Institute for Nuclear Research, 141980 Dubna, Russia} \\
$^3${\it\small Lebedev Physical Institute, Russian Academy of Sciences,}\\
{\it\small Leninsky Prospekt 53, 119991 Moscow, Russia}}

\date{ }

\begin{document}

\maketitle

\begin{abstract}
The measurement of positron--nucleus collisions was used to estimate the possibility 
of suppressing background events that simulate $\beta\beta$ decay in the emulsion 
region adjacent to molybdenum conglomerates. The range of the escape of two 
relativistic particles from the interaction was found to be 
$<d> = (0.60\pm 0.03) ~\mu$m, which approximately 
corresponds to the grain size of developed nuclear emulsion. No correlation of 
the values of d with the angle between two relativistic particles was observed. 
It was shown that it was possible to exclude $\beta\beta$ decay background from 
electrons emerging in the decay of elements of naturally occurring radioactive 
chains. The background from $\beta$ decays of $^{90}$Sr and $^{40}$K available in 
emulsion around Mo conglomerates was determined by the ratio of the volume 
$(\sim d^3)$ to the total volume of emulsion and was found to be $1.5\cdot 10^{-2}$. 
It was shown that the backgrounds from $^{40}$K, $^{90}$Sr and natural 
radioactivity could be significantly suppressed and would not limit 
the sensitivity of the experiment with 1 kg $^{100}$Mo.
\end{abstract}



Nuclear emulsion as a detector of electrons emerging in $\beta\beta$ decay has been 
used in a number of experiments \cite{FRE52,BAR}. The papers \cite{BAR} present the 
result of an emulsion experiment to search for $\beta\beta$ decay of $^{96}$Zr, 
which obtained the best (at that time) limit on the $2\nu\beta\beta$ decay of 
$^{96}$Zr. Using emulsion chambers layered with $\beta\beta$ decay source foils was 
discussed in \cite{DRA08}. We proposed an experiment \cite{ASH10} which made use of 
a nuclear emulsion with molybdenum ($^{100}$Mo) filling for $\beta\beta$ decay 
observation. The main advantage of this technique is the visualization of events and 
the possibility to measure all decay characteristics: total energy, energy of every 
electron and a decay angle between electrons. 

The energy of electrons is determined by their path range from the site of escape 
from Mo conglomerate up to the complete stop, where the path has a characteristic 
form in emulsion. In our experiment, fine powder (Mo, $2-4 ~\mu$m) was mixed with 
nuclear emulsion during its preparation. Details of preliminary tests and 
estimates of expected results of using nuclear emulsions with $^{100}$Mo filling 
were published in \cite{ASH10}. 

\begin{figure}
\begin{center}
\includegraphics[width=5cm]{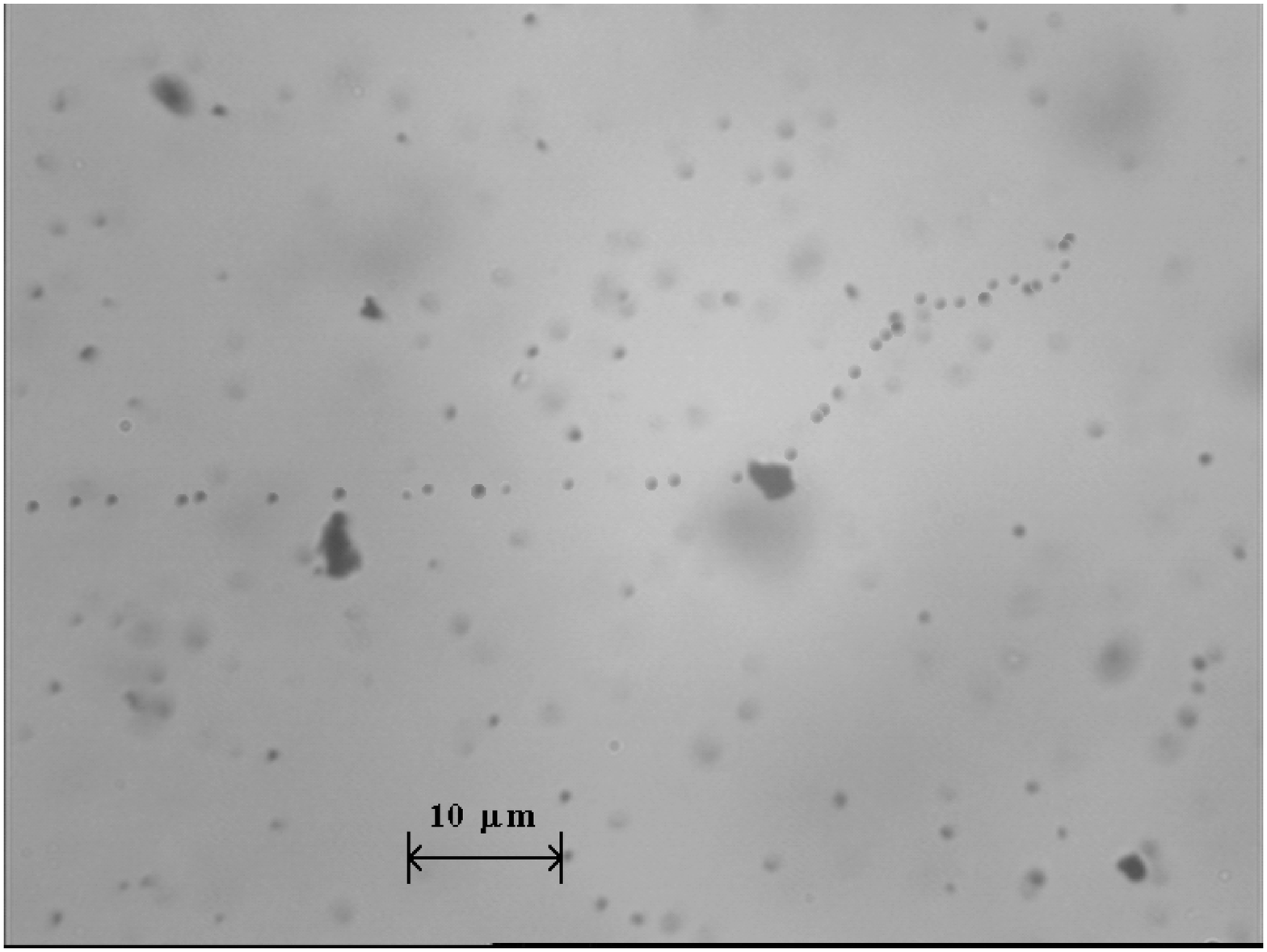}
\includegraphics[width=5cm]{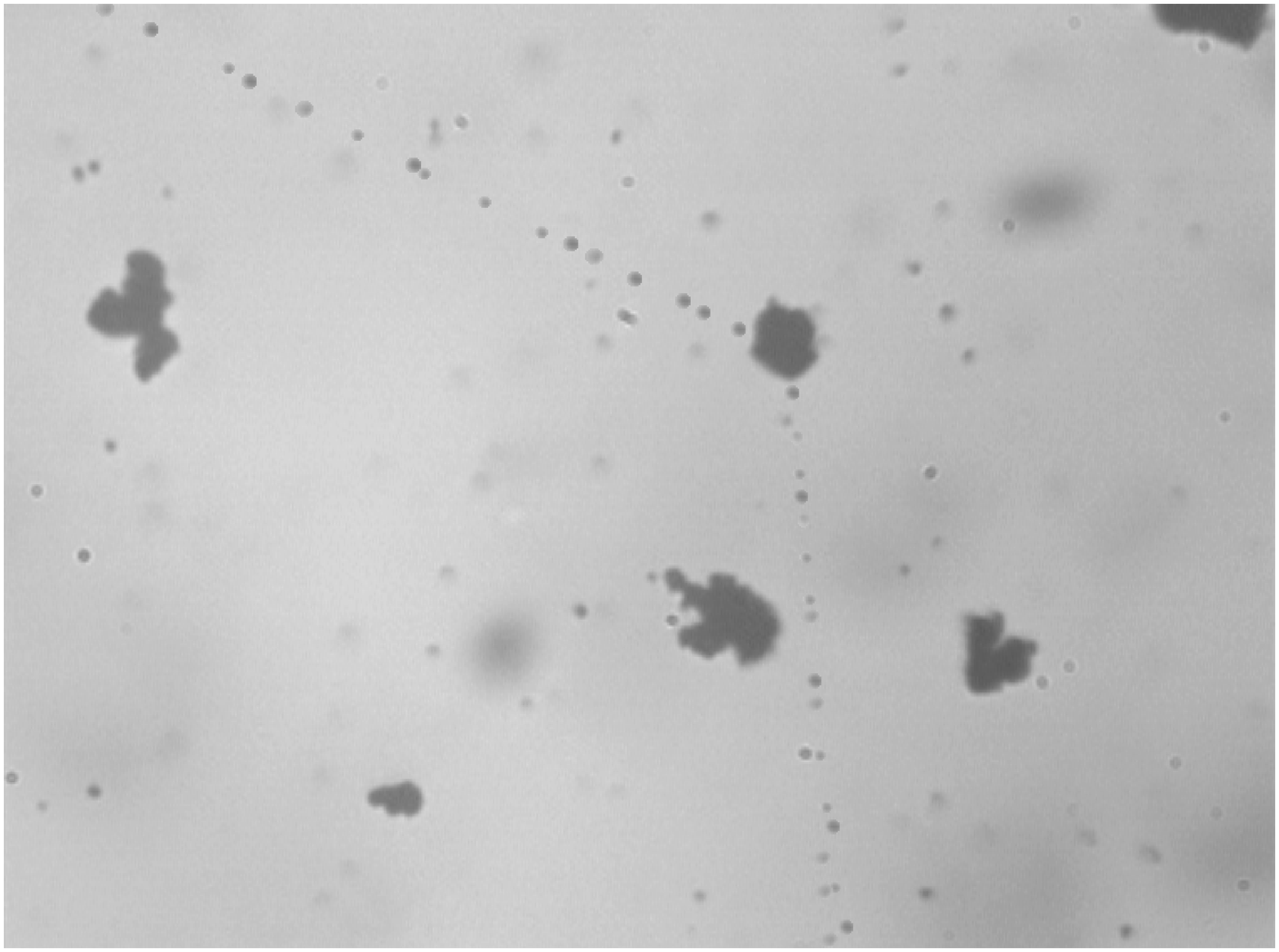}
\caption{Simulated escape of two electrons (positioned in the focal plane of the objective lens) from a Mo conglomerate.}
\end{center}
\label{fig:fig1}
\end{figure}

The present work provides the results of assessing the possibility of excluding 
some backgrounds that simulate $\beta\beta$ decay in the configuration of the 
experiment proposed in \cite{ASH10}. Figure 1 shows a real micrograph of 
conglomerates (stuck-together grains) of fine-grain commercial Mo powder in nuclear 
emulsion and a simulation of the escape of two electrons with various energies from 
Mo. If these two electrons escape as the result of $\beta\beta$ decay of the 
$^{100}$Mo nucleus, they should escape from one point. Several other factors make 
the observation of the escape of two electrons possible. 

$\bullet$ As the exposure of emulsion chambers with $^{100}$Mo is assumed to be 
prolonged, and nuclear emulsion has no temporal resolution, a successive not 
simultaneous escape of two electrons from a conglomerate would be registered as a 
candidate for a $\beta\beta$ decay of $^{100}$Mo. 

$\bullet$ Such events can be $\beta$ decays of various isotopes occurring in Mo as 
impurities due to the insufficient purification of $^{100}$Mo, and a $\beta$ decay 
of $^{40}$K ($T_{1/2} = 1.28\cdot 10^9$ years; $Q_{\beta} = 1.312$ MeV), which is 
present in gelatine near the conglomerate. The endpoint of the $\beta$ spectrum of 
$^{40}$K is 1.31 MeV. That is, the maximum energy of two $\beta$ events (from two  different decays of $^{40}$K) can reach 2.62 MeV. At not so good an energy 
resolution such events may with some (small) probability simulate events from 
neutrinoless double beta decay of $^{100}$Mo. 

$\bullet$ Background events can occur in the decay of $^{90}$Sr close to a Mo 
conglomerate. Strontium decays to yttrium ($^{90}$Sr$_{38}\to ^{90}$Y$_{39} + e^- + 
\tilde \nu_{e} (T_{1/2} = 28.8$ years; $Q_{\beta} = 0.549$ MeV)), which, in turn, 
rapidly decays to (stable) zirconium ($^{90}$Y$_{39}\to ^{90}$Zr$_{40} + e^- + 
\tilde \nu_{e} (Q_{\beta} = 2.28$ MeV)). This chain of two successive decays will 
look in emulsion as a $\beta\beta$ decay. The maximum possible energy of two 
electrons of this chain is 2.829 MeV, whereas for $^{100}$Mo it is 3 MeV. In the 
case of an insufficient energy resolution these events can contribute to the effect.

\begin{figure}
\begin{center}
\includegraphics[width=7cm]{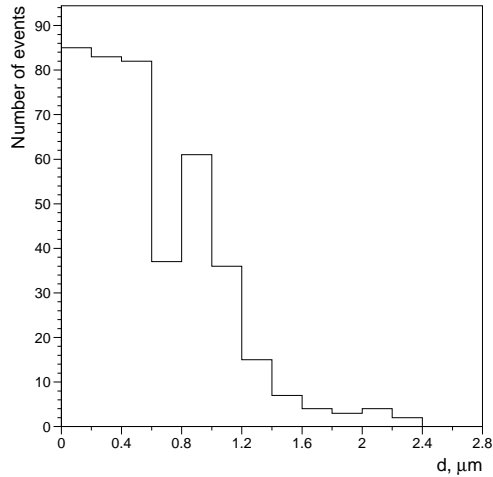}
\caption{Distribution of the value of d, a minimum distance between two tracks 
formed by particles escaping from the collision point of a positron and an emulsion 
nucleus.}
\label{fig:fig2}
\end{center}
\end{figure}

To assess the possibility of excluding $\beta\beta$ decay-simulating events 
we used positron-nucleus interactions, where relativistic particles have escaped 
from the collision point at different angles. In those measurements we arbitrarily chose pairs of relativistic particles with an angle $\phi$ between them, and
determined how exactly they intersected in the region of the interaction vertex.
The calculations used the spatial coordinates of two emulsion grains on each track,
i.e. the nearest to the interaction vertex and then the next up to the fourth grain. 
The length of the measuring base for electrons is limited due to their low energy,
and, as a consequence, strong scattering. The effect of scattering on
the accuracy of determining the intersection of electron tracks is estimated by 
the data of \cite{POM69}. Due to errors in the measurement of the coordinates of
grains on particle tracks and to scattering the straight lines drawn through these points do not intersect but are skew. Therefore, a minimal distance between 
these lines is taken to be the "intersection point" d. Figure 2 shows the distribution of the values of d. The mean value of $<d> = (0.60\pm 0.03)  ~\mu$m 
and its spatial position within the limits of angle $\phi$ does not exceed the size
of the conglomerate, and 80\% of the values of d is concentrated in the region 
of $\sim 1 ~\mu$m relative to the collision point. Figure 3 shows the dependence of
d on angle $\phi$. As seen from the data of Figs. 2 and 3, $<d>$  has the size of
approximately one developed grain of nuclear emulsion, and no correlation between d
 and angle $\phi$ is observed.

\begin{figure}
\begin{center}
\includegraphics[width=6cm]{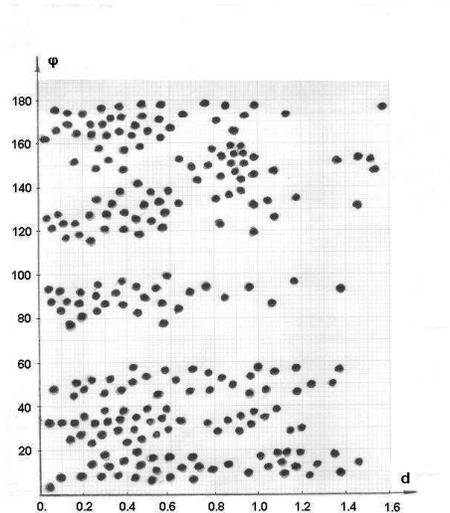}
\caption{Dependence of the value of d on the angle $\phi $ between two tracks.}
\label{fig:fig3}
\end{center}
\end{figure}

For assessing the background suppression we considered Mo conglomerates to have a
round shape, $<R_{congl}> \sim 3 ~\mu$m, and the "dangerous" zone around them 
to have the size of $\sim $ d $ (0.6 ~\mu$m). In this case, the number of $^{40}$K
 and $^{90}$Sr decays in the "dangerous" zone near all conglomerates for exposure with 1 kg of $^{100}$Mo (5.6 litres of emulsion, $n_{congl}\approx 10^{12}$) will 
be suppressed by a factor of $\sim 1.5\cdot 10^{-2}$ of the total number of 
decays in emulsion. A real number of background events depends on the content of 
potassium and strontium in gelatine. After the purification of potassium in 
gelatine up to $\sim 10^{-8}$ g/g the number of $^{40}$K decays in the "dangerous" 
zone will be $\sim 0.7\cdot 10^{-5}$ decays/year per conglomerate, and the 
probability of observing two electrons, $\sim 5\cdot 10^{-11}$. This value should 
be further reduced to account for the probability of the escape of two electrons 
from one point (a region equal to $<d>$), $\sim 0.1$. As the result, the number 
of $\beta\beta$ “events” due to $^{40}$K will be $\sim 5$ per year of exposure with 
1 kg $^{100}$Mo. Much less than 1 event will get into the energy range of 
$(3\pm 0.3)$ MeV. A more profound purification of gelatine from impurities is 
possible.

In the case of strontium, both electrons escape from one vertex, and this event may 
simulate a $\beta\beta$ decay. About $10^3$ decays will occur in the "dangerous" 
zone at a $^{90}$Sr activity in gelatine at a level of 1 mBq/kg per year of 
measurements with 1 kg of $^{100}$Mo. The contribution of two-electron events to the 
energy range of $>2.8$ MeV will be less than 1 event (at an energy resolution 
of 10\%). Besides, it should be borne in mind that in this case we would deal with "asymmetric" events (maximal energies of electrons would be $\sim 0.6$ and 
$\sim 2.2$ MeV). This can also be used to reduce the background from $^{90}$Sr, 
because in a neutrinoless double beta decay (a widespread mechanism) the most 
probable distribution of electron energy in a pair is "symmetric" (see the 
discussion of a similar situation in \cite{ARN03}).

\begin{figure}
\begin{center}
\includegraphics[width=5cm]{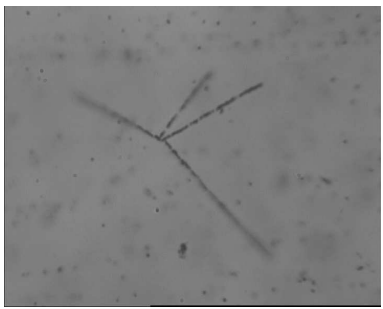}
\caption{A "star" observed in emulsion in successive $\alpha $ decays of thorium-series elements.} 
\label{fig:fig5}
\end{center}
\end{figure}

$\bullet$ Nuclear emulsion always has impurities of natural radioactive elements of 
the thorium, uranium (radium) and actinium series. In a typical (not extreme) case, 
1 cm$^3$ emulsion contains $\sim 20$ decays forming $(3-5)$-ray stars consisting 
of $\alpha$ particles (1 atom of thorium per $\sim 10^8$ atoms of emulsion) 
\cite{POW59}. It would seem that electron decays of these elements occurring in 
gelatine near Mo conglomerates may form a background which would simulate a 
$\beta\beta$ decay of $^{100}$Mo. What does occur in reality, however, is that 
a chain of decays of the three series of natural radioactivity always begins with 
the successive emission of $(3-5) ~\alpha$ particles, and only after that elements 
which are sources of $\beta$ and $\gamma$ radiation occur. In emulsion with a high 
efficiency, $3-5$-ray $\alpha$ stars are observed (Fig. 4) and even the diffusion 
of radon (Rn), which emerges in the radium series, is measured. The work 
\cite{POW59} presents the spectra of $\alpha$ particle path lengths in nuclear 
emulsion, which emerge in the decay of the nuclei of natural radioactive elements. 
These path lengths of $10-50 ~\mu$m ($\alpha$ energy, $3-9$ MeV) are well 
observed in emulsion and the efficiency of their registration is $\approx 100\%$. 
Such "double" events (a Mo conglomerate and an $\alpha$ star) make it possible to 
determine and exclude a background electron with a high probability. If a thorium 
star proves inside a conglomerate, it can be observed too, because the path range of 
 $\alpha$ particles considerably exceeds the size of Mo conglomerate. 

The work was partially supported by the RFBR grant No 11-02-00476.

 --------------------------------------------------------------


\begin{thebibliography}{References}
\bibitem{FRE52}
J.H. Fremlin and M.C. Walters, Proc. Phys. Soc., 65 (1952) 911.
\bibitem{BAR}
A.S. Barabash et al., ITEP Preprint No 13, 1987. \\
A.S. Barabash et al., ITEP Preprint No 104-88, 1988.\\
A.S. Barabash et al., ITEP Preprint No 131-90, 1990.
\bibitem{DRA08}
M. Dracos, J. Soc. Photogr. Sci. Technol. Japan, 71 (2008) 335.
\bibitem{ASH10}
V.D. Ashitkov et al., Nucl. Instr. Methods A, 621 (2010) 701.
\bibitem{POM69}
A. A. Pomansky, LPI Preprint  No 7, 1969.
\bibitem{ARN03}
R. Arnold et al., Nucl. Instr. Methods A, 503 (2003) 649.
\bibitem{POW59}
C.F. Powell, P.H. Fowler and D.H. Perkins, The Study of Elementary Particles by 
the Photographic Method, Pergamon Press: London – New York – Paris – Los Angeles, 1959.

\end{thebibliography}
\end{document}